\documentclass[twocolumn,prd,aps,showpacs,amsmath,amssymb,raggedbottom,nofootinbib]{revtex4-1}
\usepackage{graphics}
\usepackage{graphicx}
\usepackage{dcolumn}
\usepackage{bm}
\usepackage{mathrsfs}
\usepackage{slashed}
\usepackage{amsmath}
\usepackage{epsfig}
\usepackage{amsfonts}
\usepackage{amssymb}
\def\beq{\begin{equation}}
\def\eeq{\end{equation}}
\def\bea{\begin{eqnarray}}
\def\eea{\end{eqnarray}}
\def\nn{\nonumber}

\def\k{\mathbf{k}}

\begin{document}

\title{Thermal Radiation from a Fluctuating Event Horizon }
\author{E. Arias$^1$}
\email{enrike@cbpf.br}
\author{G. Krein$^2$}
\email{gkrein@ift.unesp.br}
\author{G. Menezes$^2$}
\email{gsm@ift.unesp.br}
\author{N. F. Svaiter$^1$}
\email{nfuxsvai@cbpf.br}
\affiliation{$^1$Centro Brasileiro de Pesquisas F\'{\i}sicas, Rua Dr. Xavier
Sigaud 150, 22290-180  Rio de Janeiro, RJ, Brazil\\
$^2$Instituto de F\'\i sica Te\'orica, Universidade Estadual Paulista\\
Rua Dr. Bento Teobaldo Ferraz 271 - Bloco II, 01140-070  S\~ao Paulo, SP, Brazil}

\begin{abstract}
We consider a pointlike two-level system undergoing uniformly accelerated motion.
We evaluate the transition probability for a finite time interval
of this system coupled to a massless scalar field near a fluctuating event horizon.
Horizon fluctuations are modeled using a random noise which generates 
light-cone fluctuations. We study the case of centered, stationary and Gaussian 
random processes. The transition probability of the system is obtained from the 
positive-frequency Wightman function calculated to one loop order in the noise averaging process. 
Our results show that the fluctuating horizon modifies the thermal radiation but 
leaves unchanged the temperature associated with the acceleration.
\end{abstract}


\maketitle

\section{Introduction}

Quantum field theory in curved space-time~\cite{pe1,most} describes quantum 
fields propagating in a classical gravitational field background. Important processes 
described by the theory are vacuum polarization and particle creation in cosmological 
models and black-hole evaporation~\cite{pe2}. A black hole emits thermal radiation 
due to quantum effects with an effective temperature 
inversely proportional to its mass, $\beta^{-1}= 1/8\pi M$. The basic question that 
naturally arises is the following: is there a way to measure such a radiation in a 
suitable setup? A~partial answer to this question was provided by Unruh, who introduced 
the idea of studying analogous condensed matter systems which reproduce kinematical 
features of black-hole physics. Unruh showed~\cite{pe3} that the propagation of sound 
waves in an irrotational and inviscid supersonic fluid is equivalent to the propagation 
of scalar waves in a black-hole space-time. Since this seminal paper, the possibility 
of simulating aspects of general relativity and quantum fields in curved space-time 
through analog models has been widely discussed in the 
literature~\cite{pe41,pe42,pe43,pe44,pe45,pe46,pe47,pe5,pe6}. 
In particular, the possibility of constructing an acoustic analog of a black hole and 
measuring sound waves with thermal spectrum can provide an experimental verification of the existence of 
Hawking radiation in a condensed matter setting. In this respect, there exist 
interesting recent proposals to generate an acoustic metric with sonic horizon in atomic 
Bose-Einstein condensates and other superfluids~\cite{pe71,pe72,pe73,pe74,pe75,pe76,pe77,pe78}. 

There is, however, a serious difficulty in the semiclassical picture underlying the
derivation of the thermal spectrum of back-hole radiation. Tracing the Hawking radiation 
back in time, one has to undo an exponentially strong gravitational red-shift in the 
vicinity of the horizon. This so-called trans-Planckian problem can spoil the derivation 
of the Hawking effect. The gravitational back reaction also raises questions on the 
applicability of the semiclassical theory of gravity. Within this perspective, models 
were formulated~\cite{pe81,pe82} with the aim of studying the effects of fluctuations 
of the black-hole horizon on the Hawking radiation spectrum. In the absence of knowledge 
of the precise nature of the metric fluctuations near the horizon, the assumptions made here are
that they can be treated classically and their effects on the propagation of quantum 
fields can be described via random differential equations. We emphasize that this is 
quite different from the stochastic gravity program~\cite{sgp}, where the 
Einstein-Langevin equation enables one to find the dynamics of metric fluctuations 
generated by the fluctuations on the stress tensor of quantum fields.

Once established the analogy between black holes and fluids, it is a plausible approach to 
treat random media as models for fluctuations of the effective geometry of a sonic 
black hole. In such a scenario, recently an analog model for quantum gravity effects was proposed~\cite{pe12}. 
The model builds on the work of Ford and collaborators~\cite{pe9,pe101,pe102,pe103,pe104} and 
Hu and Shiokawa~\cite{pe11}. Two general features of waves propagating in random 
fluids underlie the model. First, acoustic perturbations in a fluid define discontinuity 
surfaces that provide a causal structure with sound cones. Second, propagation of 
acoustic excitations in random media are generally described by wave equations with 
random speed of sound~\cite{pe13,pe14,pe15,pe16}. In Ref.~\cite{pe12}, the 
quantum field theory of a scalar field associated with acoustic waves was analyzed 
in a situation where the speed of propagation of the acoustic wave, and hence the sound
cone, fluctuates. A stochastic ensemble of fluctuating geometries was assumed in
that work. 

In the present paper we take a pragmatic point of view towards seeking experimental 
consequences of quantum gravity effects in a specific scenario. Specifically, 
we analyze the question whether a quantum device can detect such effects by
considering how fluctuations of a black-hole event horizon affect the transition 
rate of a two-level system which interacts with a massless scalar field. Since we are not 
assuming the rotating wave approximation, the two-level system measures the vacuum 
noise in its world line. 

The simplest assumption one can make for modeling the event horizon fluctuations is 
to assume a wave equation with random coefficients. The differential equation governing
the random wave propagation cannot be solved in closed form, but it can be treated in 
perturbation theory~\cite{pe21}, using as a small expansion parameter the intensity 
of the noise correlation function. Using such a perturbative expansion, the wave 
multi-scattering processes can be interpreted in terms of Feynman diagrams. The 
positive-frequency Wightman function is calculated at the one loop level of the noise 
averaging process. With this result in hand, it is possible to calculate the distortion 
in transition probabilities of a two-level system induced by the fluctuating horizon. 

Although the horizon fluctuations do not invalidate the
semiclassical derivation of the Hawking effect,
we show that predictions
provided by the radiative processes in the semiclassical theory scenario
may differ from those of this quantum
gravity effect scenario, where
the quantum fluctuations of the metric are treated using a stochastic ensemble of geometries. In this context, we would like 
to call the attention of the reader to the fact that Hu and Roura~\cite{roura} discussed 
a few years ago the possibility of studying the positive-frequency two-point Wightman 
function in the presence of metric fluctuations and the response function of a detector 
coupled to the field. Also, a deviation from the thermal spectrum was found by Takahashi 
and Soda~\cite{soda} using a different model for a fluctuating black-hole horizon.

The organization of the paper is as follows.
In Section II we discuss quantum field theory in the presence of a Schwarzschild event horizon.
We use the fact that, close to the horizon the Schwarzschild metric takes the form of the
Rindler line element. Note that the Rindler's line element
is static, and consequently there is a straightforward way
to define positive and negative frequency modes in order
to impose the canonical quantization in Rindler's spacetime.
With these considerations, we describe an apparatus device which is sensitive to fluctuations of the event horizon.
In Section III we
discuss quantum fields in fluctuating disordered medium.
In Section IV we present
the distortion caused by the fluctuating horizon
in the transition probabilities.
Also, in this section, using topological arguments  we show that
fluctuations in the horizon do not change the temperature
associated with the acceleration, but only the spectrum of the thermal radiation.
Finally, section V contains our conclusions. To simplify presentation we assume units 
such that $G=\hbar=c=k_{B}=1$.

\section{The Unruh-DeWitt detector}

Our aim is to discuss a particular model for fluctuations of the black-hole event 
horizon. We are interested to know how such fluctuations can affect the thermal 
radiation due to the presence of the event horizon. Therefore, let us consider the 
line element of a four-dimensional Schwarzschild space-time which describes a 
non-rotating uncharged black hole of mass $M$:
\begin{equation}
ds^2=\biggl(1-\frac{2M}{r}\biggr)dt^2-\biggl(1-\frac{2M}{r}\biggr)^{-1} dr^2
-r^{2}d\Omega^{\,2},
\label{det1}
\end{equation}
where $d\Omega^{\,2}$ is the metric of a unit $2$-sphere.
Close to the horizon, $r\approx 2M$. Therefore Eq.~(\ref{det1}) can be written 
as
\begin{equation}
ds^{2}=\biggl(\frac{\rho}{4M}\biggr)^2\,dt^{2}-d\rho^{2}-4M^2d\Omega^{\,2},
\label{det2}
\end{equation}
where $\rho(r)=\sqrt{8M(r-2M)}$. In these coordinates the horizon is at $\rho =0$.
The quantity $4M^2d\Omega^{\,2}$ describes the
line element of a  $2$-sphere of radius $2M$.
The other two terms can be identified with
the line element of the two-dimensional Rindler edge by setting
$t=4Ma\tau$ and $\rho = e^{a\xi}/a$, for $0<\rho<\infty$ and $-\infty<t<\infty$. 
Then:
\begin{equation}
ds^{2}=e^{2\,a\xi}(d\tau^{2}-d\xi^{2})-4Md\Omega^{\,2}.
\end{equation}
The null asymptotes $\xi\rightarrow\,-\infty$, $\tau\rightarrow\pm\infty$ act 
as event horizons. Note also that lines of constant $\xi$ are hyperbolae, hence they 
represent the world lines of uniformly accelerated observers. One sees that, close to 
event horizon the Schwarzschild metric takes the form of the Rindler line element. 
Therefore, in order to capture the essential physical features of such a situation, 
we consider an uniformly accelerated pointlike two-level system in a Minkowski space-time with 
light-cone random fluctuations. 

Although out of the scope of the present publication to answer the important 
question ``what is a detector and what is the phenomenon of detecting particles", 
it nevertheless requires some discussion. It is a known fact that there is a conceptual 
problem in quantum field theory in the construction of the Hilbert space of particles 
when particles are observer-dependent. The situation can be clarified to some extent 
by considering the response of an accelerated detector. In the context of quantum 
optics,  Glauber~\cite{glau}, Loudon~\cite{lou}, and Nussenzveig~\cite{nuss} proposed 
ideal photocounter detectors. Afterwards, Unruh~\cite{unruhdet} and DeWitt~\cite{dewittdet} 
proposed scalar-particle detector models. Crudely speaking, the so-called ``Unruh-DeWitt 
detector" is a two-level system with nonzero matrix 
elements of a monopole operator. The detector has the feature that its response to 
an interaction with a scalar field in the Minkowski vacuum depends on its state of 
motion. When in inertial motion, the detector has a vanishing asymptotic probability to wind 
up in an excited state, while if it moves with a constant proper acceleration, it 
has a finite asymptotic probability to undergo transition to an excited state. 
Moreover, the accelerated detector, with 
proper acceleration $\alpha$, interacting with the scalar field in the Minkowski vacuum
is equivalent to the situation of the detector in inertial motion but in contact  
with a bath of thermal radiation at the temperature $\beta^{-1} = \alpha/{2\pi}$. 
Following these results, many papers appeared in the literature studying such a 
detector in many different situations. For the reader interested in more details, 
we recommend Ref.~\cite{nb} (and references therein), in which the authors 
studied the Unruh-DeWitt detector in a situation where the probability of its 
excitation is evaluated over a finite time interval.

After this brief digression, let us describe our idealized model. In this paper 
we consider an Unruh-DeWitt detector; a two-level system coupled via a monopole 
interaction with a massless scalar field. We will be working in four-dimensional 
Minkowski space-time, whose line element is given by:
\begin{equation}
ds^{2}=dt^{2}-dx^{2}-dy^{2}-dz^{2}.
\label{det3}
\end{equation}
In order to find the distortion caused by the fluctuating event horizon
in the decay and excitation rates of a quantum system, let us discuss the
response function of a two-level system. Let $|\,g\,\rangle$ and $|\,e\,\rangle$
be energy eigenstates of the system, with eigenenergies $\omega_{g}$ and $\omega_{e}$,
and $E = \omega_{e} - \omega_{g}$ the gap between the two states. Next, we 
suppose that the system is weakly coupled to a real scalar field with interaction 
Lagrangian~\cite{ nb,det1}:
\begin{equation}
L_{int}=c \, m(\tau) \, \varphi(x(\tau)),
\label{nova1}
\end{equation}
where $x^{\mu}(\tau)$  is the world line of the two-level system
parametrized by the proper time $\tau$, and $m(\tau)$ is
the monopole-moment operator of the two-level system.
The quantity $c$ is a small coupling constant between the detector and 
the scalar field. It is clear that this is an oversimplified model, with 
an atom represented by a two-level system that interacts with a real 
massless scalar field. However, this model contains all the properties needed 
to understand the basic features of radiative processes of atoms near a 
fluctuating event horizon. We can define an initial state at $\tau=0$ given by
$|g\,\rangle\otimes|0\,\rangle$ and a final state $|e\,\rangle\otimes|\psi_{f}\,
\rangle$, at time $\tau$. Here , where $|0\rangle$ and $|\psi_f\rangle$ are the 
vacuum and final states of the field. In first order approximation 
perturbation theory in the monopole coupling constant~$c$, Eq.~(\ref{nova1}), 
the transition probability is given by:
\begin{equation}
P(\tau,0)=c^{2} \, |\langle e\,|\,m(0)|\, g \rangle|^{2} \, F(E,\tau),
\label{nova111}
\end{equation}
where $c^{2}|\langle e\,|\,m(0)|\, g \rangle|^{2}$ is the selectivity of the 
two-level system, and $F(E,\tau)$ is the response function:
\begin{eqnarray}
F(E,\tau) &=&\int_{0}^{\tau}d\tau' \int_{0}^{\tau}d\tau''\,e^{-iE(\tau'-\tau'')}
\nonumber \\ && \times\,\langle\,0|\varphi(x(\tau'))\varphi(x(\tau''))|0\,\rangle .
\label{dett1}
\end{eqnarray}
Using the definition of the positive-frequency Wightman function:
\beq
G^{\,+}(x,x') = \langle\,0|\varphi(x)\varphi(x')|0\,\rangle ,
\eeq
and introducing the variables $\zeta=\tau'-\tau''$  and $\eta = \tau'+\tau''$, 
the response function can be rewritten as:
\begin{equation}
F(E,\tau) = \frac{1}{2}\,\int_{-\tau}^{\tau}d\zeta \, e^{-iE\zeta}  
\int_{|\zeta|}^{2\tau-|\zeta|}\,d\eta \, G^{\,+}(\zeta, \eta).
\label{det10}
\end{equation}
For a free massless scalar field, one has $G^{\,+}(\zeta, \eta) = G^{\,+}(\zeta)$, 
and letting $\tau\rightarrow\infty$, the double integration would reduce to a Fourier 
transform of the Wightman function, times an infinite time integral. 
Similarly, we shall show later on in the paper that random fluctuations of the light cone introduce 
an unbounded function $f(\tau)$ that depends on the functional form of the noise
correlation function. However, the transition probability per unit proper time should be finite. 
Such circumstances often arise in quantum field theory and may be dealt 
with by adiabatically switching off the coupling as $\tau\rightarrow\pm\infty$. 
To circumvent such a problematic situation, we assume the field-detector interaction 
occurring during a finite time interval and, because of this, we choose to evaluate 
the response function over a finite proper time interval. One should recall that the 
detector defined above responds to the vacuum fluctuations because we do not assume 
the rotating wave approximation. The two-level system is measuring the vacuum noise 
in its world line. Consequently $F(E,\tau)$ defines the spectrum of the vacuum noise. 
Determination of $F(E,\tau)$ requires the positive frequency Wightman function 
$G^+(x,x')$. In the next Section we determine $G^+(x,x')$ and discuss the consequences 
of a disordered medium on the response function $F(E,\tau)$.

\section{Perturbation theory in a disordered medium}

 In order to access the modification caused by the fluctuating event horizon on transition 
probabilities, we implement a perturbation calculation similar to the one used 
in the context of problems of fluctuating disordered media and discussed in Refs.~\cite{pe12,pe21}.
Let us consider the random massless scalar Klein-Gordon equation in a four 
dimensional space-time, given by:
\begin{equation}
\left[\left(1+\,\mu({\bf r})\right) \frac{\partial^{2}}{\partial\,t^{2}}
- \Delta\right] \varphi(t,{\bf r})=0,
\label{nami20}
\end{equation}
where $\Delta$ is the three dimensional Laplacian. Note that $\mu({\bf r})$ works 
like a local refractive index, in that it perturbs locally the wave speed due to 
the replacement:   
\beq
\frac{\partial^{2}}{\partial t^{2}}\rightarrow (1+\mu({\bf r}))
\frac{\partial^{2}}{\partial t^{2}}.
\label{random-speed}
\eeq
For the random function $\mu({\bf r}\,)$ we will take a zero-mean Gaussian random 
function:
\begin{equation}
\langle \mu({\bf r}\,) \rangle_{\mu} = 0, \label{nami21}
\end{equation}
with a white-noise correlation function given by:
\begin{equation}
\langle \mu({\bf r}\,)\mu({\bf r}\,') \rangle_{\mu} =
\sigma^2\,\delta^{(3)}({\bf r}-{\bf r}\,'), \label{nami22}
\end{equation}
where $\sigma^2$ gives the intensity of random fluctuations.
The symbol $\langle\, ...\,\rangle_{\mu}$ denotes an average over
all possible realization of this random variable. 
On the other hand, in principle it is possible to extend the method to colored and/or non-Gaussian noise
functions. Note that we assume a time-independent random function for inertial observers. However, in the 
accelerated detector world line $\mu$ becomes a Rindler time-dependent random function.
We will discuss this point later on in the paper.

Following Refs.~\cite{pe12,pe21}, the random Klein-Gordon equation of 
Eq.~(\ref{nami20}) can be solved using a perturbation expansion in the
noise function. In this way, the positive- frequency Wightman function can 
be written as: 
\begin{eqnarray}
\hspace{-0.25cm}
G^{\,+}(x,x') &=& G^{\,+}_{0}(x-x') \nonumber\\ 
&& +
\sum_{n=1}^{\infty}\int dz_1 \,G^{\,+}_{0}(x-z_1){\cal G}^{(n)}(z_1,x'),
\label{a1}
\end{eqnarray}
where $G^{\,+}_{0}(x-x')$ is the usual positive-frequency Wightman function 
without random fluctuations, and
\begin{equation}
{\cal G}^{(n)}(z_1,x') = (-1)^{n}\prod_{j=1}^n L_1(z_j)
\int dz_{j+1} \,G^{\,+}_{0}(z_j,z_{j+1}),
\label{genericterm}
\end{equation}
with $L_1 (x)$ being the random differential operator:
\beq
L_1 (x) = L_1(t,{\bf r\,}) = - \mu({\bf r\,}) \frac{\partial^2}{\partial t^2}.
\label{L1-coord}
\eeq
In Eq.~(\ref{genericterm}), it is to be understood that $z_{n+1} = x'$ and 
that there is no integration in $z_{n+1}$. Details on the derivations of 
the above expressions can be found in Ref.~\cite{pe21}. 

Due to the Gaussian nature of the noise averaging, higher order correlation functions 
of the form $\langle \mu({\bf r}_{1}) \mu({\bf r}_{2}) \cdots \mu({\bf r}_{p}) \rangle_{\mu}$ 
can be easily expressed as the sum of products of two-point correlation functions corresponding 
to all possible partitions of ${\bf r}_{1},{\bf r}_{2}, \cdots , {\bf r}_{p}$. 
An interesting feature of wave propagation in random media is Anderson 
localization~\cite{Anderson}. In this context, we note that truncation of the 
series in Eq.~(\ref{genericterm}) at a finite order $n$ will miss the singular 
aspect of the localization problem, which is of a non-perturbative nature --
see Refs.~\cite{pe18,pe181,pe182,pe183,pe184,pe185,pe186,pe187} for discussions 
on this and related subjects. For our purposes in the present paper it is 
sufficient to use only the first terms of the series. 

After performing the averages over the noise function, the connected 
two-point positive-frequency Wightman function associated with the massless scalar 
field can be written in the form of a Dyson equation:
\begin{equation}
\langle G^{\,+} \rangle_{\mu} =  G^{\,+}_{0} - G^{\,+}_{0}
\Sigma \, \langle G^{\,+} \rangle_{\mu},
\label{p32}
\end{equation}
with $\Sigma$ being the self-energy. The one-loop contribution to $\Sigma$ is
obtained from noise averaging the second order contribution in the random 
function $\mu$. Specifically, one can write:
\begin{equation}
\langle \, G^{\,+}(x,x') \rangle_{\mu}= G^{\,+}_{0}(x-x') 
+ \langle\bar G_{1}(x,x')\rangle_{\mu},
\label{gg}
\end{equation}
and going over to Fourier space:
\begin{equation}
\langle\bar G_{1}(x,x')\rangle_{\mu} = \int\frac{d{\bf k}}{(2\pi)^3}\int\frac{d\omega}{(2\pi)} \,
e^{-ik(x-x')} \, \langle\bar G_{1}(\omega,\k)\rangle_{\mu},
\label{FT-G1}
\end{equation}
allows us to write:
\begin{eqnarray}
\langle\bar G_{1}(\omega,\k)\rangle_{\mu} &=& - G^{\,+}_{0}(\omega,\k) \Sigma(\omega) G^{\,+}_{0}(\omega,\k),
\label{eu}
\end{eqnarray}
with $G^{\,+}_{0}(\omega,\k)$ being:
\beq
G^{\,+}_{0}(\omega,{\bf k})=i/(\omega^2-{\bf k}^2),
\label{G0}
\eeq
and
\begin{equation}
\Sigma(\omega) = - \sigma^2\omega^4\, \alpha(\omega) ,
\label{Sigma}
\end{equation}
where the quantity $\alpha(\omega)$ is given by: 
\begin{equation}
\alpha(\omega) = \int \frac{d{\bf k}}{(2\pi)^3} \, G^{\,+}_{0}(\omega,\k) = 
-\frac{\omega}{4\pi}.
\label{alpha}
\end{equation}

\begin{figure}[h]
 \centering
 \includegraphics[scale=0.45]{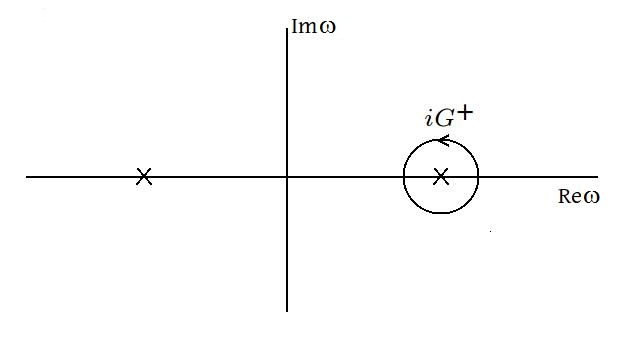}
\caption{The contour of integration in the complex $\omega$ plane for the positive-frequency Wightman function.}
\label{contornok0}
\end{figure}

Using these results in Eq.~(\ref{FT-G1}), one obtains the correction due to 
the random fluctuations to the Wightman function in terms of the integral:
\begin{eqnarray}
\langle\bar{G}_{1}(x,x')\rangle_{\mu} &=&\frac{\sigma^2}{2\,(2\pi)^5}
\int d{\bf k} \int d\omega \, e^{-i\omega(t-t')+i{\bf k}\cdot({\bf x}-{\bf x}')}
\nonumber\\ 
&& \times\,\frac{\omega^5}{\left(\omega^2-{\bf k}^2\right)^2}.
\label{G11}
\end{eqnarray}
The integral over $\omega$ can be done considering contour of integration in 
Fig.~(\ref{contornok0}). The integrand in Eq.~(\ref{G11}) has a second-order pole 
in $\omega=|{\bf k}|$. The final result, after performing the integration 
${\bf k}$, is: 
\begin{eqnarray}
\hspace{-0.5cm}
\langle \bar{G}_{1}(x,x') \rangle_{\mu} &=& \frac{i}{(2\pi)^3} \, \frac{6 \sigma^2}
{\left[(\Delta t-i\epsilon)^2 
- (\Delta{\bf x})^2\right]^5} \biggl[ (\Delta t)^5 \nn \\[0.3true cm]
&& + \, 10 \, (\Delta t)^3 \, (\Delta {\bf x})^2 
- 3 \, \Delta t (\Delta {\bf x})^4 \biggr],
\label{G1}
\end{eqnarray}
where $\Delta t=t-t'$ and $\Delta{\bf x}={\bf x}-{\bf x}'$.

Now we are in position to take this result over different trajectories of 
the Unruh-DeWitt detector. This will be done in the next Section.

\section{Transition probabilities} 

\begin{figure}[t]
 \centering
 \includegraphics[scale=0.3]{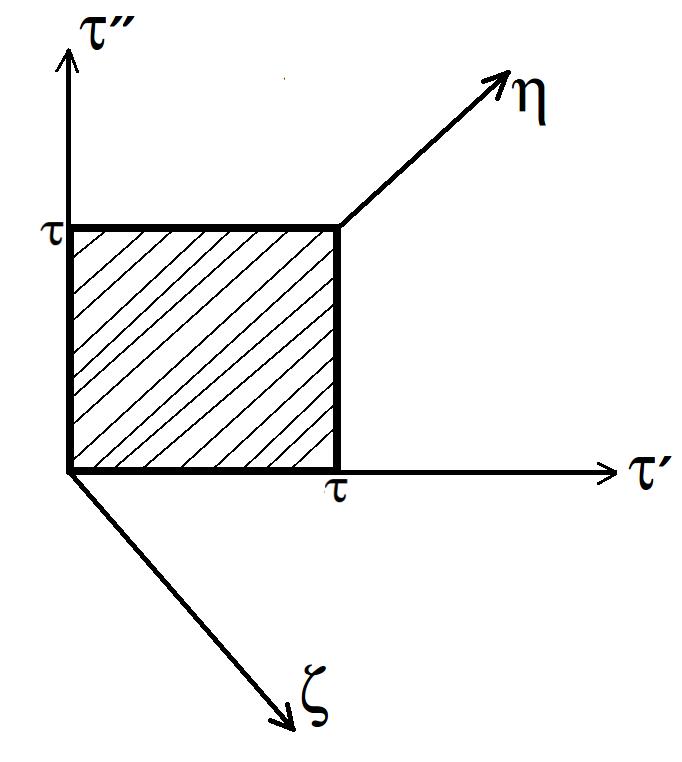}
\caption{Region of integration in the variables $\eta$ and $\zeta$.}
\label{int}
\end{figure}

Having obtained the two-point positive-frequency Wightman function 
associated with the massless scalar field, it can be used to find the 
response function of the two-level system: 
\begin{equation}
F(E,\tau)=\frac{1}{2}\,\int_{-\tau}^{\tau}d\zeta\,e^{-iE\zeta}
 \int_{|\zeta|}^{2\tau-|\zeta|}\,d\eta\,
 \langle G^{\,+}(\zeta, \eta)\rangle_{\mu},
\label{dis1}
\end{equation}
where $\langle G^{\,+}(\zeta, \eta)\rangle_{\mu}$ can be obtained from 
Eq.~(\ref{gg}) and (\ref{G1}) -- Fig.~\ref{int} illustrates the coordinate 
system for the integration over $\eta$ and $\zeta$. Next, one shall consider the 
Unruh-DeWitt detector moving in $(t,x)$ plane along a hyperbolic trajectory:
\beq 
x = (t^2 + \alpha^{-2})^{1/2},\,\,\,\, y = z = 0, 
\label{hyperb}
\eeq
with $\alpha$ being a constant. As well known, this represents a detector 
accelerating uniformly with acceleration $\alpha$ in the frame of the detector. 
The detector's proper time $\nu$ is related to $(t,x)$ by the relations:
\begin{equation}
t = \alpha^{-1}\sinh(\alpha \nu) ,
\label{t}
\end{equation}
and
\begin{equation}
x = \alpha^{-1}\cosh(\alpha \nu) .
\label{x}
\end{equation}

Performing such a coordinate transformation in Eq.~(\ref{gg}), one
can write the response function to one-loop order as a sum of 
three contributions:
\begin{equation}
 F(E,\tau) = F_{\beta}(E,\tau) + F_0(E,\tau) + F_1(E,\tau) .
\label{dis7}
\end{equation}
The first term is the usual thermal contribution and the second 
is due to the switching on and off of the coupling between the 
two-level system and the scalar field. We are interested in the third 
term since it contains the correction due
to fluctuating event horizon. The second term vanishes when 
large time intervals are considered. Using the techniques developed 
in Ref.~\cite{nb}, the first term in Eq.~(\ref{dis7}) reads:
\begin{eqnarray}
F_{\beta}(E,\tau) &=& \frac{\tau\,|E|}{2\pi}  \Biggl[\Theta(-E)\biggl( 1 + 
\frac{1}{e^{ \beta |E|} - 1}
\biggr) \nonumber\\  && + \, \Theta(E) \; \frac{1}{e^{ \beta E }-1}\Biggr] ,
\label{Fbeta}
\end{eqnarray}
where $\beta = 2\pi/\alpha$. The transition rate, i.e. the probability of 
spontaneous and induced decay and excitation per unit time, of the two-level 
system given by:
\beq
R_\beta(E) = \frac{d}{d\tau}F_\beta(E,\tau),
\label{rate}
\eeq
can be readily computed. Clearly, the result of Eq.~(\ref{Fbeta}) is that one has 
the same effect that of a bath of thermal radiation at a temperature $\beta^{-1} 
= \alpha/2\pi$ -- see Refs.~\cite{unruhdet,davies}. Within the perspective of 
the thermalization theorem, the result can be put in the following form: the pure 
state which is the vacuum from the point of view of an inertial observer is a 
canonical ensemble from the point of view of a uniformly accelerated observer, 
with a temperature proportional to the magnitude of the observer's acceleration.

The contribution $F_1(E,\tau)$ due to the fluctuating event horizon is given by
\begin{equation}
F_1(E,\tau)= \frac{1}{2}\,\int_{-\tau}^{\tau}d\zeta\, e^{-iE\zeta}
\int_{|\zeta|}^{2\tau-|\zeta|}\,d\eta\, \langle\bar{G}_{1}(\zeta,\eta)\rangle_{\mu},
\label{ult}
\end{equation}
and can be evaluated in the same way as $F_\beta(E,\tau)$. Performing the coordinate 
transformations given in Eqs. (\ref{t}) and (\ref{x}) in Eq.~(\ref{G1}), 
one finds:
\begin{equation}
\langle\bar{G}_{1}(\zeta,\eta)\rangle_{\mu} = \frac{6i}{(2\pi)^3}
\bigg(\frac{\alpha}{2}\bigg)^5
\frac{\tilde{\sigma}^2(\eta)}{\sinh^5\bigg(\alpha\zeta/2 
- i \epsilon\alpha\bigg)},
\label{G1-eta-zeta}
\end{equation}
where we have absorbed a positive function of $\eta$ and $\zeta$ into the infinitesimal parameter $\epsilon$. Also, the quantity $\tilde{\sigma}^2(\eta)$ that gives the 
intensity of the horizon fluctuations and is given by:
\begin{eqnarray}
\tilde{\sigma}^2(\eta) &=& \sigma^2\cosh\biggl(\frac{5\,\alpha\,\eta}{2}\biggr).
\label{mod}
\end{eqnarray}
The $\eta$ dependence comes from the fact that in the Wightman function, $x$ 
is not an independent variable; rather, it is determined by the detector's trajectory. 
In other words, inertial observers in this model experience static light-cone random fluctuations -- see Eq.~(\ref{nami20}). However, for uniformly accelerated observers, 
such fluctuations will not be static anymore and will also depend on their proper 
times. On physical grounds, it is clear that for an uniformly accelerated detector 
the effects of the fluctuations will increase with its proper time, a result that 
is manifest in Eq.~(\ref{mod}). 

\begin{figure}[t]
 \centering
 \includegraphics[scale=0.3]{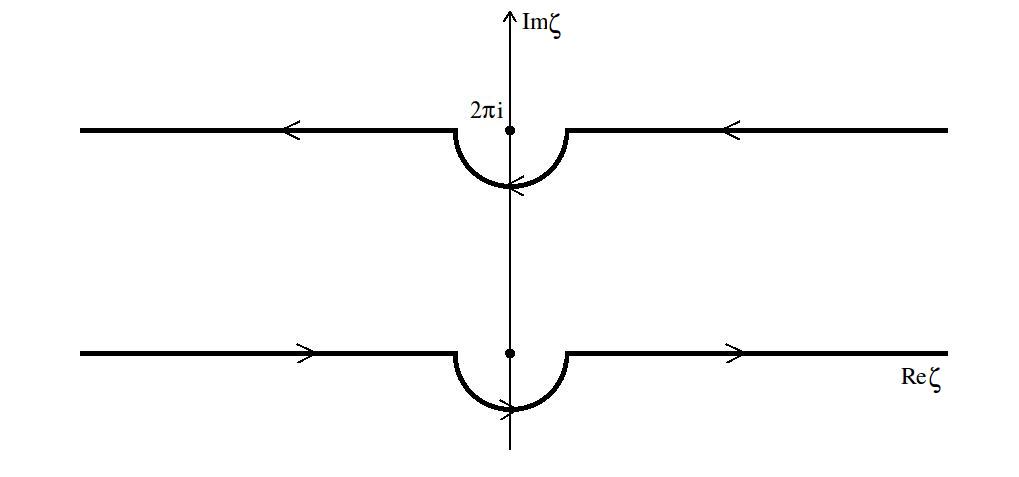}
\caption{The contour of integration appropriate to the evaluation of integral $I$.}
\label{contorno}
\end{figure}

Using the result of Eqs.~(\ref{G1-eta-zeta}) and (\ref{mod}) in Eq.~(\ref{ult}), 
one obtains the expression:
\begin{eqnarray}
F_1 (E,\tau) &=& \frac{6 i}{5(2\pi)^3} \bigg(\frac{\alpha}{2}\bigg)^4 \, \tilde{\sigma}^2(\tau) \, \int_{-\tau}^{\tau} d\zeta \, e^{-iE\zeta}  \nonumber \\[0.3true cm]
&& \times \, \frac {\sinh [5\alpha (\tau - |\zeta|)/2] }
{\sinh^5 (\alpha\zeta/2 - i\epsilon\alpha)}.
\label{integral}
\end{eqnarray}
We may express the integral above as 
\begin{eqnarray}
\int_{-\tau}^{\tau}\,d\zeta f(\zeta) &=& \int_{-\infty}^{\infty}\,d\zeta f(\zeta) 
\nonumber \\[0.3true cm]
&& - \, \left[\int_{-\infty}^{-\tau}\,d\zeta f(\zeta)\, + \,\int_{\tau}^{\infty}\,d\zeta f(\zeta) \right], 
\label{trick}
\end{eqnarray}
where
\begin{equation}
f(\zeta)=
\frac{1}{\alpha} \, e^{-iE\zeta/\alpha} \; 
\frac{ \sinh \left[5(\alpha\tau - \,|\zeta|)/2 \right]}{\sinh^5(\zeta/2)}.
\end{equation}
The last two-terms on the right-hand side of Eq.~(\ref{trick}) can be expressed 
as a single integral. They have the same physical origin as the term $F_0(E,\tau)$. 
In order to perform the integral
\begin{equation}
I = \int_{-\infty}^{\infty}\,d\zeta f(\zeta),
\end{equation}
we may use contour integration -- the contour to be used is shown in
Fig.~(\ref{contorno}). The integral over the lower part of the contour yields 
$I$ while that over the upper part yields $\exp{(2\pi E/\alpha)}I$. The sum of these contributions is related to the fifth-order residue of $f(\zeta)$
at $\zeta=0$. Finally, collecting these results, we have that
\beq
F_1(E,\tau) = W(E,\tau) + H(E,\tau), 
\label{F1-sum}
\eeq
where
%
\begin{equation}
W(E,\tau) = -\frac{\tau\tilde{\sigma}^2(\tau)}{(4\pi)^2}\frac{\big(24\alpha^4 - 35 E^2\alpha^2+E^4\big)}{e^{2\pi E/\alpha}+1},
\label{R2noinercial}
\end{equation}
and
%
%
\begin{eqnarray}
H(E,\tau) &=& \frac{12\tilde{\sigma}^2(\tau)}{5(2\pi)^3}
\left(\frac{\alpha}{2}\right)^4 \int_{\tau}^{\infty} d\zeta \, \sin(E\zeta) 
\nn \\[0.3true cm]
&& \times \,\frac{\sinh \left[5\alpha(\tau - |\zeta|)/2 \right]}
{\sinh^5 \left[\alpha\zeta/2 \right]}.
\label{R2noinercial2}
\end{eqnarray}
%
In Eq. (\ref{R2noinercial}) we consider $\tau$ as a small quantity, so that $\sinh(5\alpha\tau/2) \approx 
{5\alpha\tau}/{2}$ .

Eqs~(\ref{F1-sum})-(\ref{R2noinercial2}) comprise our main result in this paper. 
We have found that the correction due to horizon fluctuations gives a thermal 
distribution with a temperature $\beta^{-1}=\alpha/2\pi$, that is the same 
temperature as for the non-fluctuating case, {\em but} the distribution is of 
Fermi-Dirac form. This resembles the result obtained by Takagi~\cite{takagi}, 
who studied the power spectrum of the vacuum noise measured by an accelerated 
detector in arbitrary dimensions and discussed the phenomenon of inversion of 
statistics in odd dimensions. Nevertheless, as emphasized by the literature~\cite{unruh86}, this is an 
apparent inversion of statistics. In our case we have a Fermi-Dirac correction to 
the thermal radiation of a bosonic field, but we still have the expected 
Bose-Eintein distribution as the leading component of the radiation. The 
meaning of our result is that horizon fluctuations imply in a radiation 
spectrum of both contributions, in that the usual Bose-Einstein distribution
is perturbed by a Fermi-Dirac distribution of the same temperature. 
One could expect that higher loop corrections would give additional 
energy-dependent terms that will be neither Fermi-Dirac nor 
Bose-Einstein forms. 

It is easy to see from our result how the fluctuating horizon will 
change the transition rate of a detector undergoing an inertial 
trajectory by taking the zero proper acceleration limit in 
Eq.~(\ref{R2noinercial}):
\begin{equation}
 \lim_{\alpha\rightarrow0}  F_1(E,\tau) = - \tau\frac{\sigma^2}{16\pi^2}E^4
 \Theta(-E)
\end{equation}
In the case of an inertial detector the intensity of the fluctuations will not change with time.
The correction to the transition rate for the inertial detector is proportional to
$\Theta(-E)$, meaning that for an inertial detector we have only spontaneous decay 
induced by the vacuum fluctuations.

Let us study the dependence of the temperature with random fluctuations in more detail.
We will be using arguments developed by Christensen and Duff \cite{duff}.
The Euclidean manifold associated to inertial observers has the topology of
${\cal{R}}^{4}$, with Euler-Poincar\'e characteristic $\chi=1$. For the case of the
uniformly accelerated observer, we have a non-simply connected manifold, with
topology ${\cal{R}}^{3}\times{\cal{S}}^{1}$, with Euler-Poincar\'e characteristic $\chi=0$.
Now if we have these two distinct topological situations, we can define two
different vacua, one associated with inertial observers, defined by $|\chi=1\rangle$ and another one associated with accelerated observers, defined by $|\chi=0\rangle$. Next, 
one can show that an accelerating observer regards the $|\chi=1\rangle$ vacuum as a 
thermal state at temperature $\beta^{-1} = \alpha/{2\pi}$. The two-point Schwinger 
function associated with the massless scalar field for both cases can be defined 
as:
\begin{equation}
G^{(\chi)}(x,x')=\langle\,\chi|\varphi(x)\varphi(x')|\chi\,\rangle.
\label{ult1}
\end{equation}
This two-point function obeys
\begin{equation}
\Delta G^{(\chi)}(x,x')=-\delta^{4}(x-x'),
\label{ult2}
\end{equation}
where $\Delta$ is now the four dimensional Laplacian.  For simplicity, let us take
the case where the two points $x$ and $x'$ belong to the accelerated world-line
$\xi=\xi'=\alpha^{-1}$, $y=y'$ and $z=z'$. The generalization for two arbitrary points
can be found in Ref.~\cite{carlos}.  We have:
\begin{equation}
G^{(1)}(\tau,\tau')=\frac{1}{16\pi^{2}}\frac{\alpha^{2}}
{\sin^{2} \bigl(\alpha \Delta \tau /{2}\bigr)},
\label{ult3}
\end{equation}
and, since $G^{(1)}(\tau,\tau')$ must be periodic in $\tau$ with period $2\pi/{\alpha}$,
we have:
\begin{equation}
G^{(1)}(\tau,\tau')=G^{(1)}(\tau,\tau'+2\pi/{\alpha}).
\label{ult4}
\end{equation}
In the case where $\chi=0$, paths winding around the origin have topologically
distinct classes with winding number~$n$. In this case we have
\begin{equation}
G^{(1)}(\tau,\tau')=\sum_{n=-\infty}^{\infty}G^{(0)}(\tau,\tau'+2\pi n/{\alpha}).
\label{ult5}
\end{equation}
Using the KMS condition, we have that the finite temperature Schwinger
function satisfies
\begin{equation}
G_{\beta}(\tau,\tau')=G_{\beta}(\tau,\tau'+\beta),
\label{ult6}
\end{equation}
and using the identification $\beta^{-1}=\alpha/{2\pi}$ we have
that $G^{(0)}=G_{\beta\rightarrow\infty}$ and  $G^{(1)}=G_{\beta}$.
We conclude that the ground sate of the accelerated observer is the $|\chi=0\rangle$ vacuum state, relative to which the  $|\chi=1\rangle$  vacuum is a thermal state. In the presence
of noise, it is easy to see that the above arguments can be used in the same way. 
Recalling Eq.~(\ref{p32}), we get:
\begin{equation}
\langle G^{(1)}(\tau,\tau')\rangle_{\mu}=\langle G^{(1)}(\tau,\tau'+2\pi/{\alpha}) \rangle_{\mu}.
\label{ult7}
\end{equation}
In the case where $\chi=0$, we also have
\begin{equation}
\langle G^{(1)}(\tau,\tau') \rangle_{\mu}=\sum_{n=-\infty}^{\infty}\langle G^{(0)}(\tau,\tau'+2\pi n/{\alpha}) \rangle_{\mu}.
\label{ult8}
\end{equation}
Using the KMS condition, we have that the finite temperature Schwinger
function satisfies
\begin{equation}
\langle G_{\beta}(\tau,\tau') \rangle_{\mu}=\langle G_{\beta}(\tau,\tau'+\beta) \rangle_{\mu},
\label{ult9}
\end{equation}
and using the same identification $\beta^{-1} = \alpha/{2\pi}$ we have
that $\langle G^{(0)} \rangle_{\mu}=\langle G_{\beta\rightarrow\infty} \rangle_{\mu}$ and
$\langle G^{(1)} \rangle_{\mu}=\langle G_{\beta} \rangle_{\mu}$.
Therefore, the temperature associated with the acceleration
remains the same, but the fluctuating horizon does change its emitted thermal 
radiation.


\section{Conclusions}

Recently, an analog model for quantum gravity effects in a condensed matter 
scenario was proposed in Ref.~\cite{pe12}. In Ref.~\cite{pe21} this discussion 
was extended to a more general case, namely to a massive real scalar field. 
In the present paper such a model was
used to study a massless scalar field near a four-dimensional 
Schwarzschild black hole with fluctuations in the event horizon. Using a 
perturbation theory similar to the one used in problems of fluctuating disordered 
media, we obtain the two-point positive-frequency Wightman function 
associated with a real scalar field. After performing the averages over 
the noise function, we discuss the thermal radiation near the fluctuating 
event horizon. We obtained the modification of the transition probabilities 
caused by the fluctuating horizon on the decay and excitation processes.
We showed that horizon fluctuations imply that the usual Bose-Einstein distribution
is perturbed by a Fermi-Dirac distribution of the same temperature. 
Our results are obtained by assuming that the 
mechanical disturbances caused by the radiative processes belongs to an energy scale much smaller than the
mass of the accelerated two-level system. This means that its world line does not change due to light-cone random fluctuations.  
We conclude our discussions noting that the previous treatment can
be presented using the Fermi Golden Rule and the fact that the density
of states per unit volume is given by $-{1}/{\pi} \, {\rm Im}G_{R}({\bf r},{\bf r}\,')$,
where $G_{R}$ is the retarded Green function associated with the massless scalar field.

Finally, we remark that random-matrix theory
can be used to find how the spectral density near the
fluctuating event horizon is modified. In a similar situation,
in Ref. \cite{benaker1} the transition rate of a two-level atom within 
a chaotic cavity was presented using random matrices.
The use of random matrices to find how the spectral density near the
fluctuating event horizon is modified is under investigation by the
authors


\acknowledgments
N. F. Svaiter would like to acknowledge the hospitality of the
Instituto de F\'{\i}sica Te\'{o}rica, Universidade Estadual
Paulista, where part of this research was carried out. G. Menezes would like 
to acknowledge the hospitality of the Physics and Astronomy Department of 
Tufts University, where part of this research was carried out.
We would like to thank L.H. Ford, E. Goulart and F. F. Tovar 
for useful discussions. This paper was supported by CAPES, CNPq and FAPESP (Brazilian
agencies).

\end{document}